\documentclass[global,twocolumn]{svjour}
\usepackage{latexsym}
\usepackage{graphicx}

\journalname{myjournal}
\newcommand{\beq}{\begin{equation}}
\newcommand{\eeq}{\end{equation}}
\newcommand{\beqa}{\begin{eqnarray}}
\newcommand{\eeqa}{\end{eqnarray}}
\newcommand{\nnb}{\nonumber}
\newcommand{\reflef}{(\ref}

\newcommand{\bcent}{\begin{center}}
\newcommand{\ecent}{\end{center}}
\begin{document}
\title{Probing the semi-macroscopic vacuum by higher-harmonic
       generation under focused intense laser fields}

\author{K.~Homma\inst{1}${}^,$\inst{2}, D.~Habs\inst{2} and T.~Tajima\inst{2}}

\offprints{Kensuke Homma}
\institute{
$^1$ Graduate School of Science, Hiroshima University, Kagamiyama, 
Higashi-Hiroshima 739-8526, Japan\\
$^2$ Fakult\"at f\"ur Physik, Ludwig Maximilians 
Universit\"at M\"unchen, D-85748 Garching, Germany
}
\date{Received: date / Revised version: date}

\maketitle
\begin{abstract}
The invention of the laser immediately enabled the detection of
nonlinear photon-matter interactions, 
as manifested for example by Franken et al.'s
detection of second-harmonic generation.
With the recent advancement in high-power, high-energy lasers
and the examples of nonlinearity studies of the laser-matter interaction
by virtue of properly arranging lasers and detectors,
we envision the possibility of probing nonlinearities of the photon
interaction in vacuum over substantial space-time scales, compared to the
microscopic scale provided by high-energy accelerators.
Specifically, we introduce the photon-photon interaction in a quasi-parallel
colliding system and the detection of higher harmonics in that system.
The method proposed should realize a far greater sensitivity
of probing possible low-mass and weakly coupling fields 
that have been postulated.
With the availability of a large number of coherent photons, 
we suggest a scheme for the detection of higher harmonics
via the averaged resonant production and decay of these postulated fields
within the uncertainty of the center-of-mass energy
between incoming laser photons.
The method carves out a substantial swath of new experimental parameter
regimes on the coupling of these fields to photons,
under appropriate laser technologies,
even weaker than that of gravity in the mass range well below 1~eV.
\end{abstract}

\sloppy
\section{Introduction}
  \label{sec1}
%
%
Recent astronomical observations suggest the existence of non-luminous stuff
in the universe: dark matter and dark energy. 
Understanding their origin is one of the greatest scientific challenges of the
21st century. It may be natural to expect that even 
the universe is just a collection of vacua around us.
Therefore, we consider a novel method to directly probe 
the structure of vacuum under laboratory conditions that relate to these issues.
To our current knowledge no direct signature of the non-luminous stuff
has ever been observed in terrestrial experiments.
This is in spite of the fact that numerous advanced theories exist, including 
axions~\cite{PDG}, minicharged particles~\cite{MCP0,MCP1,MCP2,MCP3},
and dark energy~\cite{DEreview,FujiiScalarTensor}.
The reason for this may be thought of as follows.
After Rutherford's discovery of the inner core (i.e. nucleus) of an atom
being very tiny compared to the dimension of the already tiny size of the atom,
the experimental search went to explore ever smaller constituents of matter and
thus the thrust went for higher-energy or momentum experiments.
Theories have gone hand-in-hand
with this exploration, succeeding in ever shorter-ranged interaction
theories and unification of forces, as exemplified by the electroweak 
theory~\cite{Weinberg}. We refer to this standard and extremely successful method
as the high-momentum approach. Almost all laboratory
research efforts have been on this approach to date.
Although successful in exploring high energy physics, this approach is not 
suitable to explore energies much lower than 1~eV.
These fields, that might exist in much lower
domains than 1~eV, cannot strongly couple to matter,
because if they did, they would have been long ago
discovered in the low-energy region. Thus these fields, if they ever exist, must
couple weakly. This means that we need an extremely strong driver to
manifest a sufficiently strong signal overcoming this weakly coupling
interaction, showing up above noise.
So far such sufficiently powerful photon sources did not exist. 
However, this may be changing now with more intense
lasers becoming available~\cite{MourouRMP}.
What we have called the high-amplitude or high-field
method~\cite{TajimaHighFieldScience,TajimaEPJD}
may provide an alternative path to detect such low-mass, weakly
interacting fields that are spread over semi-macroscopic scales.

When the energy of the constituent is much lower than that of the probing 
photon, conceptually we may employ two laser beams in a co-propagating 
geometry. The two collinear beams produce a very low center-of-mass 
energy interaction at the beating frequency 
(being equal to the difference of the two laser frequencies)~\cite{Beat}.
This interaction could resonate with the very low eigenfrequency of the
constituent, should there be an eigenmode in its vicinity.
Second of all, the co-propagating setup
allows us to make the two beams interact over a much
prolonged interaction time, thus much amplifying the nonlinearities and signals
arising from these.

\begin{figure}
\includegraphics[width=1.0\linewidth]{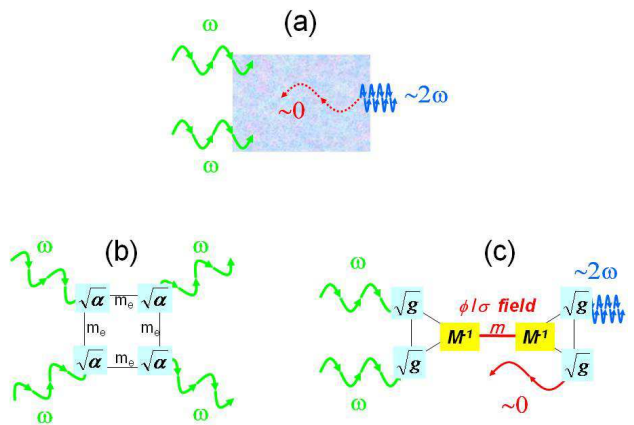}
\caption{
Schematic diagrams of photon-photon interactions in matter and in vacuum.
(a) Second harmonic generation in the experiment by Franken et al.
~\cite{Franken}, arising from the nonlinearity of crystal fields irradiated
by (intense enough) laser fields;
(b) Probing QED vacuum nonlinearities as suggested by
Heisenberg and Euler~\cite{EH,Weiscop},
where the vertices of the coupling in the Feynman diagram are characterized
by the fine structure constant of the vacuum $\alpha$.
Thus a weak nonlinearity requires much more intense
fields than in the case (a). The leading order interaction is the elastic
photon-photon scattering, though as a higher order there exists a second
harmonic generation as well; (c) Probing potential light-mass $m$ fields
in vacuum with intense laser fields. The vertices are characterized by very
feeble couplings of $M^{-1}$ and $g$~\cite{FujiiScalarTensor}.
The expected second-harmonic generation may be said
to be not different to case (a). In order to increase
the observable signal, suggestions have been made.
}
\label{Fig6}
\end{figure}

In order to pick up the experimental signal of a strong coupling to
the long-range mode, we suggest using higher-harmonic
generation.
The pioneering research by Franken et al.~\cite{Franken}
detected the nonlinearity in a quartz crystal via second-harmonic generation.
Two photons in co-propagation accentuate the interaction 
through the quartz fields
over the coherence volume in order to initiate second-harmonic generation.
This process may be schematically looked upon as the case displayed 
in Fig.~\ref{Fig6}(a).
There the quartz nonlinearities mix two forward-propagating photons
$(\omega)$ to produce a photon with $2\omega$ (and possibly another photon
with frequency $\sim 0$, also referred to as optical rectification
or difference frequency mixing).
The quantum electrodynamics (QED) process is illustrated by
Fig.~\ref{Fig6}(b). Two incoming photons are mediated
by virtual electron-positron fields and outgoing are two photons.
The extreme forward-scattering amplitude with quasi-parallel incident photons
is known to be largely suppressed in the QED process, as we discuss later.
This is because the center-of-mass energy of the colliding two photons
is too low to satisfy the relevant mass scale of the electron-positron pair.

Given these hints from analogous pictures on how to probe medium-like features
of vacuum, this paper rather considers a simple photon-photon scattering 
process via the averaged resonant production and decay 
of light-mass fields within the energy uncertainty in the
center of mass system between incoming laser photons.
This gives a solid basis for the design of experiments based 
on intense co-propagating lasers, as much independent as possible 
of models on the light-mass fields in vacuum.
As we discuss later in detail, the process we focus on is based on
Fig.~\ref{Fig6}(c).

%
%
The QED process as illustrated in Fig.~\ref{Fig6}(b) is described by
the Euler-Heisenberg effective Lagrangian~\cite{EH} in the low-frequency limit 
of two incoming photons. This predicts that the vacuum under the influence of an 
electromagnetic field induces a birefringence characteristic. 
The ratio between the first and the second term 
in the Lagrangian can yield a general test to see whether the vacuum 
contains other effects beyond QED, examining its value at 4:7. In general,
a scalar field $\phi$ and a pseudoscalar field $\sigma$ in vacuum
may contribute to the first and second term, respectively.
Light scalar fields as candidates of
dark energy have been recently intensively discussed~\cite{DEreview},
while the pseudoscalar fields (axion-like-particles) may be a source
of dark matter~\cite{PDG} and also possibly dark energy~\cite{deVega}.
We suggested how to measure the phase shift of a probe laser
across an intense electromagnetic field,
based on the phase-contrast Fourier imaging~\cite{apb-qed}.
However, we recognize that the test by the phase-contrast Fourier imaging
is rather limited in the mass and coupling of those fields~\cite{qed-limit}.
Therefore, we extend our method to search for those new
types of fields by instituting co-propagating laser beams.
This approach may be looked upon as in Fig.~\ref{Fig6}(c).
Again two parallel photons come in, while two parallel photons come out.
As we discuss later, the frequency shift in vacuum is simply explained
by the strong Lorentz boost in the quasi-parallel colliding system, which
causes the blue- and red-shift of photon energies emitted
to the forward and backward directions, respectively.
We note that our approach is similar to, but distinct from
many laboratory experiments with lasers~\cite{BFRT,PVLAS,BMV,ALPS,LIPSS,OSQAR,GammeV}
already performed and proposed to search for those low-mass fields.

%
%
Consider more details of the effective interaction Lagrangian $L$
as illustrated in the triangle part in Fig.~\ref{Fig6}(c), where 
the scalar field $\phi$ and the pseudoscalar field $\sigma$ couple
to the electromagnetic field via quantum anomaly-type
couplings~\cite{FujiiScalarTensor}
\beqa\label{eq_phisigma}
-L_{\phi}=g_{\phi} M_{\phi}^{-1} \frac{1}{4}F_{\mu\nu}F^{\mu\nu} \phi,
\\
-L_{\sigma}=g_{\sigma} M_{\sigma}^{-1}
\frac{1}{4}F_{\mu\nu}\tilde{F}^{\mu\nu} \sigma,
\label{mxelm_1}
\eeqa
where $F_{\mu\nu} = \partial A_{\mu} / \partial x^{\nu} -
                    \partial A_{\nu} / \partial x^{\mu}$ is
the antisymmetric field strength tensor with the 
four-vector potential $A_{\mu}$ of the electromagnetic wave and its dual tensor
$\tilde{F}^{\mu\nu} = 1/2 \epsilon^{\mu\nu\iota\kappa} F_{\iota\kappa}$
with the Levi-Civita symbol $\epsilon^{\mu\nu\iota\kappa}$.
The difference $\phi$ versus $\sigma$ is in their allowed couplings to
polarization states of two photons as we discuss later.
Here $gM^{-1}$ provides the coupling strength.
$g$ and $M$ carry the subscripts $\phi$ and $\sigma$, respectively,
indicating the corresponding type of fields.
The dimensionless coupling $g$ is typically proportional to the dimensionless
fine structure constant $\alpha$ for the two photons to couple to the
virtual charged-particle pair in the triangle part. The effective coupling
includes the large mass scale $M$ to couple to the light field
with a mass $m$.
The large $M$ induces the weakness of the coupling via $M^{-1}$.
For example, the Newtonian constant $G$ is expressed as
$8\pi G = \hbar c M^{-2}_P$, where $M_P$ is the Planckian mass of $10^{27}$~eV.
The weakness of $G$ is the manifestation of the large mass scale at
the vertex in the triangle coupling.
In what follows we omit the subscripts $\phi$ and $\sigma$ on
the coupling $gM^{-1}$ and the mass of the light field $m$,
unless we need to explicitly distinguish the type of the fields.
We use the natural units $\hbar = c = 1$
throughout the subsequent sections, unless explicitly noted.

%
%
As a quite challenging case we have attempted a theoretical approach 
to search for an extremely light scalar field as a candidate of dark energy
via the averaged resonance scattering process in \cite{DEsearch}.
Given intense laser fields in the near future,
the method may provide a new window into scoping physics on the
Planckian mass scale by photon interactions in a quasi-parallel
incident laser beam in the laboratory.
In this paper we review the essence of the approach and further
develop basic formulae in Sect.~\ref{sec2} to apply the method 
to a specific experimental setup considered in Sect.~\ref{sec3},
in order to discuss reachable limits on the coupling
strength $gM^{-1}$ and the mass $m$ for a laser intensity
attainable within the ELI project~\cite{ELI} in Sect.~\ref{sec4}.
In Sect.~\ref{sec5} we provide a perspective on
how our method serves as a new type of scope to probe the 
semi-macroscopic vacuum, which has not been extensively investigated yet.

\section{Quasi-parallel photon-photon interaction in low-mass and
         weakly coupled fields}
  \label{sec2}
%
%
\subsection{Kinematics of quasi-parallel system}
\begin{figure}
\includegraphics[width=1.0\linewidth]{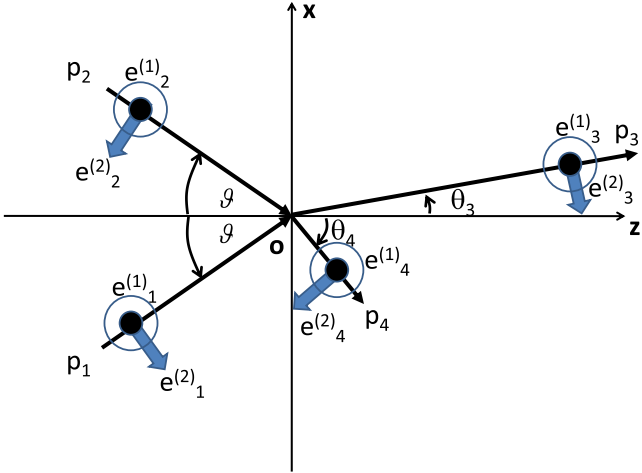}
\caption{
Definitions of kinematical variables for the suggested co-propagating
photons (this figure is quoted from \cite{DEsearch}).
}
\label{Fig7}
\end{figure}

As illustrated in Fig.~\ref{Fig7},
we introduce an unconventional coordinate
frame, in which two photons labeled by 1 and 2 sharing the same frequency are
incident nearly parallel to each other, making an angle $\vartheta$ with
the common central line along the $z$ axis. We define the $z-x$ plane
formed by $\vec{p}_1$ and $\vec{p}_2$.
The components of the 4-momenta of the photons are given by
$p_1 =(\omega\sin\vartheta,0,\omega\cos\vartheta ; \omega)$ and
the same for $p_2$ but with the sign of $\vartheta$ reversed, and
$p_3 =(\omega_3 \sin\theta_3, 0, \omega_3 \cos\theta_3 ; \omega_3)$ and
$p_4$  with $\omega_3, \theta_3$ replaced by $\omega_4, -\theta_4$,
respectively.
The angles $\theta_3$ and $\theta_4$ are defined as shown
in Fig.~\ref{Fig7}. This coordinate system can be transformed to
the center-of-mass (CM) system for the head-on collision ($\vartheta =\pi/2$)
by the Lorentz transformation
with $v/c\rightarrow 1$ for $\vartheta \rightarrow 0$.
Conversely, this implies that the realization of the quasi-parallel collision
in the laboratory frame corresponds to the realization of an
extremely low CM energy, as we see below.

In this frame one of the final photons in the forward direction along
the $z$ axis must have an upshifted frequency due to the energy-momentum
conservation, independent of the physical origin of the dynamics.
In the limit of $\vartheta \rightarrow 0$, a process of
$\omega_3 \rightarrow 2 \omega$
is realized. This frequency doubling nature is an extremely valuable
characteristics from the experimental point of view,
as compared to the case with no frequency shift in the
center-of-mass system. In addition, more importantly,
it is essential to maintain a quasi-parallel nature of the incident
beams in order to access resonance, as we shall stress later.

The energy-momentum conservation laws requires following relations;
\beqa
0 \mbox{-axis}:&&\omega_3 +\omega_4 = 2\omega, \label{kinm_3}\\
z\mbox{-axis}:&&\omega_3 \cos\theta_3 + \omega_4 \cos\theta_4=
2\omega \cos\vartheta, \label{kinm_4}\\
x\mbox{-axis}:&&\omega_3\sin\theta_3 =\omega_4\sin\theta_4.
\label{kinm_5}
\eeqa
From the conditions $0<\omega_{3,4} <2\omega$, we may choose
$0<\theta_3<\vartheta<\theta_4<\pi$,
without loss of generality.  From Eq.~\reflef{kinm_3})-\reflef{kinm_5})
we derive the relation
\beq
\sin\theta_3 =\sin\theta_4\frac{\sin^2\vartheta}
{1-2\cos\vartheta\cos\theta_4+\cos^2\vartheta}.
\label{kinm_7}
\eeq

The differential elastic scattering cross section per solid angle
$d\Omega_3$ favoring the higher photon energy $\omega_3$ is given by
\beq
\frac{d\sigma}{d\Omega_3}=(8\pi \omega)^{-2}
\sin^{-4}{\vartheta} (\omega_3/2\omega)^2 |{\cal M}|^2,
\label{kinm_13}
\eeq
where ${\cal M}$ is the invariant amplitude and
\beq
\omega_3 =\frac{\omega \sin^2\vartheta}{1-\cos\vartheta \cos\theta_3}.
\label{kinm_8}
\eeq
Here we expect the upshifted frequency $\omega_3 \rightarrow
2\omega$, as $\theta_3\rightarrow 0$ for  $\vartheta \rightarrow 0$, as
mentioned before.

%
%
%
%
\subsection{Dynamics of two-photon interaction via resonance}
We consider the scattering amplitudes only for the case when light-mass fields
are exchanged via resonance (s-channel amplitude).
The resonance decay rate of the low-mass field with the
mass $m$ into two photons is expressed as
\beq
\Gamma=(16\pi)^{-1} \left( g M^{-1}\right)^2 m^3.
\label{mxelm_4a}
\eeq
The low-mass field is exchanged between the pairs
$(p_1, p_2)$ and $(p_3, p_4)$, thus giving the squared four-momentum
of the field
\beq
q_s^2 =\left(p_1+p_2\right)^2 =2\omega^2 \left( \cos 2\vartheta -1
\right)
\label{eq_qs}
\eeq
with the metric convention $(+++-)$ for the definition of the four momenta.
$q_s$ corresponds to the CMS energy of the photon-photon collision.

With the polarization vectors given by
$\vec{e}_i^{\ (\beta)}$, where $i=1, \cdots,4$ are the photon labels, whereas
$\beta =1,2$ are for the kind of linear polarization as depicted in
Fig.~\ref{Fig7}, we summarize the non-zero invariant amplitudes for scalar
field exchanges
\beq\label{eq_M11}
{\cal M}_{1111}={\cal M}_{2222}=-{\cal M}_{1122}=- {\cal M}_{2211},
\eeq
and for pseudoscalar field exchanges
\beq\label{eq_M12}
{\cal M}_{1212}={\cal M}_{1221}=-{\cal M}_{2112}=- {\cal M}_{2121},
\eeq
where the first two digits in the subscripts
correspond to the states of the linear polarization
of the incoming two photons 1 and 2, respectively, and the last two correspond
to those of the outgoing two photons 3 and 4, respectively,
as illustrated in Fig.~\ref{Fig7}.

We focus on one of these non-zero amplitudes by denoting it as ${\cal M}$;
\beq
{\cal M} =-(g M^{-1})^2\frac{\omega^4 \left(
\cos2\vartheta -1\right)^2}{2\omega^2 \left(
\cos2\vartheta -1\right)+m^2},
\label{mxelm_7}
\eeq
where the denominator, denoted by ${\cal D}$ in the following, 
is the low-mass field propagator.
We note that $q_s^2$ in (\ref{eq_qs}) is time-like.  
We then make the replacement
\beq
m^2 \rightarrow \left( m -i\Gamma \right)^2 \approx
m^2 -2im \Gamma.
\label{mxelm_9}
\eeq
Substituting this into the denominator in Eq.~\reflef{mxelm_7}) and
expanding around $m$, we obtain
\beq
\hspace{-.1em}{\cal D}\approx -2\left( 1-\cos2\vartheta  \right) \left( \chi+ia
\right),\quad\hspace{-.7em}\mbox{with}\quad\hspace{-.7em} \chi =\omega^2 -\omega_r^2,
\label{mxelm_10}
\eeq
where
\beq
\omega_r^2 =\frac{m^2/2}{1-\cos 2\vartheta },\quad
a=\frac{m \Gamma}{1-\cos 2\vartheta}.
\label{mxelm_12}
\eeq
From Eq.~\reflef{mxelm_4a}) and \reflef{mxelm_12}),  $a$ is also expressed as
\beq\label{eq_a}
a = \frac{\omega^2_r}{8\pi}\left(\frac{g m}{M}\right)^2,
\eeq
which explicitly shows the proportionality to $M^{-2}$.
We then finally obtain the expression for the squared amplitude as
\beq
|{\cal M}|^2 \approx  (4\pi)^2 \frac{a^2}{\chi^2+a^2}.
\label{mxelm_13}
\eeq
As for the off-resonance case $\chi\gg a$, $|{\cal M}|^2$
is largely suppressed due to the factor $a^2 \propto M^{-4}$ for
the case of a small coupling $M^{-1}$.
On the other hand, if experiments take the limit of
$\omega \rightarrow \omega_r$,
$|{\cal M}|^2 \rightarrow (4\pi)^2$ is realized from (\ref{mxelm_13}) ideally.
This is independent of the smallness of the factor $M^{-4}$,
as expected from the off-resonance case or equivalently
from the square of (\ref{mxelm_7}).
This is the most important feature arising from the resonance that overcomes
the weak coupling stemming from the large relevant mass scale such as $M=M_P$.
However, we are then confronted with an extremely narrow width $a$
for {\it e.g.} $gm \ll 1$~eV , $M \sim M_P=10^{27}$~eV and
$\omega_r \sim \mbox{1 eV}$.
We now discuss how to overcome this difficulty.

\subsection{Enhancement by averaging including resonance}
In conventional high-energy collisions, the beam momenta are implicitly
supposed to be their mean values. This is because the momentum spread,
or the uncertainty expected from the de Broglie wavelength of the
relativistic particle, is negligibly small compared to the
relevant momentum exchanges in the interaction
that experiments are interested in.
Resonance searches in such experiments must adjust $\chi$ in Eq.~(\ref{mxelm_10})
and Eq.~(\ref{mxelm_13}) such that the mean $\chi$ is close enough to the
peak location within $\pm a$.
On the other hand, in the case of co-propagating laser
beams aiming at the detection of extremely small momentum exchanges via
the resonance, the situation is quite different
due to the nature of the incident waves. This is because
the uncertainty included in the initial photon momenta
is much larger than the relevant energy scale of the resonance,
leading to the condition $|\chi| \gg a $.
In this case the squared scattering amplitude must be integrated
over the possible uncertainties on the incident
wave function on an event-by-event basis.
As we will explain below, the uncertainty in the incident photon momenta
is related to the uncertainty in $\chi$ via the uncertainty in the incident
angle $\vartheta$. Therefore, it is instructive to consider the feature
of the integral of the resonance function
of the Breit-Wigner (BW) formula~\cite{BW} from $\chi_-$ to $\chi_+$ as follows;
\beq\label{eq_IntBW}
I = \int_{\chi_-}^{\chi_+}  \frac{a^2}{\chi^2+a^2} d\chi
  = \left[a\tan\left(\frac{\chi}{a}\right)\right]_{\chi_-}^{\chi_+},
\label{mxelm_14}
\eeq
where $I=a\pi/2$ and $I=a\pi$ for $\chi_+=-\chi_-=a$ and
$\chi_+=-\chi_-=\infty$, respectively.
This indicates that the value of the integral is proportional to $a$, 
{\it i.e.}, $M^{-2}$ from Eq.~(\ref{eq_a}).
The value ranges for the finite and infinite integrals over
only a factor of two. From this fact, we expect that
the integral enhances the squared scattering amplitude by a factor of $M^2$
compared to the non-resonant interaction proportional to $M^{-4}$ from
Eq.~(\ref{mxelm_7}), as long as the peak is contained within the experimental
resolution on $\chi$, {\it i.e.}, the condition $\chi_+ > a$ and $\chi_- < -a$
is satisfied.
This implies that experiments in the co-propagating laser beam
configuration need no efforts to adjust $\chi$ close to the
extremely narrow resonance region, thanks to the subsequent huge enhancement
by the integral over the wide range on $\chi$.
Meanwhile, it is difficult to identify
the exact location of the resonance mass within the wide gate on $\chi$.
Since we are interested in having a sensitivity to an extremely weak
coupling such as gravity $M=M_P$, the enhancement of the squared amplitude
is more crucial than finding the exact location of the resonance masses.
As we discuss in the following sections, however,
we may be able to provide a crude estimate on the order of
the mass scale of the resonance even in such a situation.

The consideration above leads us to parametrize the squared scattering amplitude
as follows:
\beqa\label{eq_M2ave}
\overline{|{\cal M}|^2}=\int_{-\infty}^{+\infty} \rho(p_1, p_2)
|{\cal M}|^2 d\chi
\quad \qquad \qquad \qquad \nnb\\
= (4\pi)^2\int_{-\infty}^{+\infty}\rho(p_1, p_2)
\frac{a^2}{\chi^2(p_1,p_2)+a^2}d\chi,
\label{mxelm_15}
\eeqa
where $\rho(p_1, p_2)$ is the normalized probability distribution
to supply nominal combinations $(p_1, p_2)$ from the incident laser fields.
The real part $\chi(p_1,p_2)$ is indirectly specified by $(p_1, p_2)$
via the incident angle $\vartheta$ between the incident two photons.
This parametrization expresses averaging over
possible combinations of $p_1$ and $p_2$.
We note that $(p_1, p_2)$ are not {\it a priori} the observed momenta,
but just a nominal specification among the possible initial momenta.
In other words, $\rho(p_1, p_2)$ is not a statistical weight on the
discrete momenta after the contraction of the wavepacket of each photon state.
This implies that an infinite statistics is not necessary to obtain
the continuous nature. Rather, we need this treatment even for
a two-photon state as long as the source of photons is
not a perfect plane wave.
This allows for a continuous integral on $\chi$ via the continuous combination of
$\rho(p_1, p_2)$ in Eq.~(\ref{eq_M2ave}).
As long as the probability weight $\rho(p_1, p_2)$ is close enough to
unity around the resonance peak, the enhancement discussed
with Eq.~(\ref{eq_IntBW}) is guaranteed. This is the essence of our main
strategy of the co-propagating configuration in order to overcome
the difficulty due to the narrow resonance width $a$.

In order to design experiments, we start from the resonance condition,
the first of Eq.~\reflef{mxelm_12}), by assuming $\vartheta \ll 1$,
\beq\label{eq_CMreso}
m \sim 2\vartheta_r \omega_r
\eeq
where the subscript $r$ in both angle and energy refers to a state
satisfying a resonance condition.
We note that the product $2\vartheta_r\omega_r$, corresponds to the CM energy
of the incident two photons.
This indicates that experiments have two adjustable handles for a given
mass scale or the CM energy.
We emphasize that we can lower the CM energy by several orders
of magnitude by only introducing smaller values of $\vartheta$ with 
a fixed $\omega$.
This should be contrasted to high-energy colliders, where a large
effort is needed to increase the CM energy by an order of magnitude.
This advantage is also supported from a technical point of view,
since scanning the incident angle $\vartheta$ should be much easier than
scanning the energy $\omega$ of the resonance.
We point out that the resonance condition in Eq.~(\ref{eq_CMreso}) is not 
just given at
one point, but rather in a hyperbolic band in the $\vartheta-\omega$ plane,
with a finite resolution with $\delta\vartheta$ of 
the incident angle $\vartheta$.
This implies that the deviation $\delta\omega$ from the resonance energy
$\omega_r$ can satisfy the same resonance condition with a different 
$\vartheta$ within $\pm \delta\vartheta$.
As far as $\delta\omega/\omega_r \ll \delta\vartheta/\vartheta_r$ is satisfied
in the setup, we can ignore the effect of $\delta\omega$.
This is in fact the case, as can be seen in the following discussions.
Therefore, we can take the attitude that we fix the incident energy
at the optical frequency $\omega = \omega_{opt}$
and scan $m$ by changing $\vartheta$ around $\vartheta_r$,
where $\omega_{opt}$ and $\vartheta_r$ satisfy the resonance condition
based on Eq.~(\ref{eq_CMreso})
\beq\label{eq_omegaopt}
\omega^2_{opt} = m^2/(4\vartheta^2_r).
\eeq
From Eq.~(\ref{eq_CMreso}) and Eq.~(\ref{eq_omegaopt}) 
we see that detecting lower-mass fields requires smaller values of $\vartheta_r$,
namely the parallelism of incident photons.
In this case the variation on $\vartheta$ leads to
the variation on $\chi$. This is expressed by the following relation
based on the second part of Eq.~(\ref{mxelm_10})
\beq\label{eq_the2x}
\chi(\vartheta) = \omega^2_{opt} - \frac{m^2}{4\vartheta^2}
= \omega^2_{opt} (1-\varepsilon^{-2}),
\eeq
where
$\varepsilon \equiv \vartheta / \vartheta_r$ in the unit of $\vartheta_r$
is introduced.

We now discuss the average of the squared amplitude $\overline{|{\cal M}|^2}$
over the possible uncertainty on the incident angle $\vartheta$
\beqa\label{eq_the2eps}
\overline{|{\cal M}|^2} &=&
\int^{\pi/2}_{0} \rho(\vartheta) |{\cal M}|^2 d\vartheta
\quad \qquad \nnb\\
&=& \int^{\pi/(2\vartheta_r)}_{0} \rho(\varepsilon)
|{\cal M}|^2 \vartheta_r d\varepsilon,
\eeqa
where $\rho(\vartheta)$ is a probability distribution function
normalized between 0 and $\pi/2$
as a function of the continuous uncertainty on $\vartheta$
from arbitrarily chosen two-photon combinations within a laser pulse.
The incident angle $\vartheta$ is re-expressed with $\varepsilon$
by $d\vartheta = \vartheta_r d\varepsilon$ for the second part 
of Eq.~(\ref{eq_the2eps}).
From the relation Eq.~(\ref{eq_the2x}), we find $\varepsilon = (1-x)^{-1/2}$ and
$d\varepsilon = 1/(2\omega^2_{opt}) \varepsilon^3 dx$.
Equation (\ref{eq_the2eps}) is then re-expressed with $\chi$
\beqa\label{eq_eps2x}
\overline{|{\cal M}|^2} = (4\pi)^2 \frac{\vartheta_r}{2\omega^2_{opt}}
\qquad \qquad \qquad \qquad \qquad \qquad \nnb\\
\int^{1-(2\vartheta_r/\pi)^2}_{-\infty}
\frac{\rho((1-\chi)^{-1/2})}{(1-\chi)^{3/2}} \frac{a^2}{(\chi^2+a^2)} d\chi,
\quad
\eeqa
where Eq.~(\ref{mxelm_13}) is substituted.
This equation is the exact representation of Eq.~(\ref{eq_M2ave}), starting from
the uncertainty on the incident angle $\vartheta$, if the upper limit
of the integral range is regarded as large enough compared to $a$.
Let us define $x \equiv a\xi$ to explicitly discuss the structure of the
integral kernel in units of the width $a$ of BW.
With $\xi$, Eq.~(\ref{eq_eps2x}) is further re-expressed as
\beqa\label{eq_x2xi}
\overline{|{\cal M}|^2} =
(4\pi)^2 \frac{\vartheta_r}{2\omega^2_{opt}} a
\qquad \qquad \qquad \qquad \qquad \qquad \nnb\\
\int^{a^{-1} \{1-(2\vartheta_r/\pi)^2 \}}_{-\infty}
\frac{\rho((1-a\xi)^{-1/2})}{(1-a\xi)^{3/2}} \frac{1}{\xi^2+1} d\xi,
\eeqa
where the first factor of the integral kernel corresponds to a normalized
weight function and the second is BW with a width of unity.
This expression explicitly shows the enhancement by the factor of $a$, implying
the proportionality to $M^{-2}$ based on Eq.~(\ref{eq_a}).
As long as $\rho$ is a monotonic function, the weight function in front of
BW can be close to unity for small $\xi$ because of $a\xi \ll 1$.
With such a weight we expect that the value of the integral may be
close to that of BW, as we discuss with a specific weight function
in the following section.

The remaining issue is how to further cope in experiments with the problem 
of still very small values of $M^{-2}$, although much larger than $M^{-4}$.
First, this can be solved by the $\sin^{-4}\vartheta$ behavior of the cross
section in Eq.~(\ref{kinm_13}), that arises from the phase volume factor and
the flux factor in the quasi-parallel two-photon interaction. For an extremely
light mass, this factor gains a large number due to the small $\vartheta_r$.
Second, the intense laser fields can provide a large luminosity and
the intensity of the signal is proportional to the square of the intensity
of the laser in the limited case of an incoherent two-photon interaction.
We have three ingredients or knobs: the $M^2$ enhancement by the weighted
BW integral, the $\vartheta_r$ dependence and
the growth of the laser intensity. By marshalling these knobs,
we expect to increase the detectability for undiscovered low-mass fields
in vacuum, which have evaded from our grasp to date.

In the following sections we consider experimental realizations
with $\omega_r \sim 1$~eV (optical laser), aiming at
the mass range as low as possible.
We then plug explicit weight functions into Eq.~(\ref{eq_x2xi}),
based on the suggested experimental setup. By combining Eq.~(\ref{eq_x2xi}) and
Eq.~(\ref{kinm_13}), we discuss reachable mass-coupling limits for a
given laser intensity attainable in near future experiments.

\section{Second-harmonic detection in the Quasi-Parallel System}
   \label{sec3}

We emphasized the importance of the photon-photon interaction in a
quasi-parallel system or a small-angle setup, in order to 
enhance the signal due to low-mass constituents.
A simple way is to explore the mass range
$m<\pi\omega$ by using two independent laser beams with
a small incident angle.
We then directly measure the resonance curve in Eq.~(\ref{eq_CMreso}) by
scanning both $\vartheta$ and $\omega$ to quantitatively
observe the nature of the resonance curve. For the much smaller mass scale,
or equivalently smaller incident angle,
however, we must take into account the beam spread in the diffraction limit.
This determines the controllable smallest incident
angle, or the mass scales of the light fields, we look for.
We consider here the case of a single focused laser beam, 
in order to provide the simplest basis
to quantify reachable mass-coupling limits for a given set of experimental
parameters.
We concentrate on the detection near the second-harmonic, rather than
the laser frequency itself, to enhance the sensitivity of detecting the
photon-photon interaction.

\begin{figure}
\bcent
\includegraphics[width=1.0\linewidth]{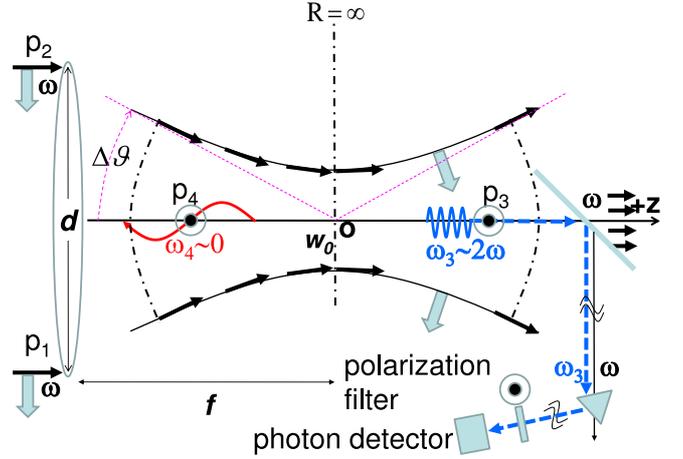}
\caption{
Suggested experimental setup for the co-propagating photon
interaction and detection. The linear polarizations of incident
and outgoing photons are drawn only for the scalar exchange
with a scattering amplitude $|{\cal M}_{1122}|$ as an example.
}
\label{Fig8}
\ecent
\end{figure}

The conceptual experimental setup with a single-beam focusing geometry
is illustrated in Fig.~\ref{Fig8}.
Incident photons from a Gaussian laser pulse with linear polarization
are focused by a ideal lens into the diffraction limit.
Quasi-parallel incident photons interact with each other
between the lens and the focal point, from which photons 3 and 4 are emitted
in nearly opposite directions along the $z$ axis
with $\omega_3\sim 2\omega$ and $\omega_4 \sim 0$.
The dichroic mirror is transparent for the
non-interacting photons with the beam energy of $\omega$,
while $\omega_3$ is reflected to the prism
(equivalent to a group of dichroic mirrors),
which selects $\omega_3$ among residual $\omega$
and sends it to the photon detector placed off the $z$-axis.
This process is assisted by a polarization filter.
From the polarization dependence of the invariant amplitude in
Eq.~(\ref{eq_M11}) and Eq.~(\ref{eq_M12}), the combinations of polarizations of
two photons between the initial and final states must satisfy
$11 \rightarrow 11(22)$ for a scalar field exchange and
$12 \rightarrow 12(21)$ for a pseudoscalar field exchange, respectively.
We note that we can choose the type of fields we search for
by setting the initial polarization state. In the case of single-beam focusing,
the search for a scalar field is easier, because we do not have to
mix the two polarization states as in the case of a pseudoscalar field.
Furthermore, the selection of the rotated final state $22$ can enhance the
signal-to-background ratio for the scalar field case, because a huge
number of non-interacting photons has the final polarization state of $11$.
In what follows we provide formulae to evaluate the accessible limit on the
mass-coupling defined in Eq.~(\ref{mxelm_1}) for a given laser intensity,
in the case that we detect a double-frequency photon per laser shot.

%
%
The Gaussian profile is a basic constraint in typical laser fields,
where the aperture of a lasing material has a finite
size in the transverse area. The solution of the electromagnetic field
propagation in vacuum with a Gaussian profile in the transverse plane
with respect to the propagation direction $z$ is well-known~\cite{Yariv}.
The electric field component in spatial coordinates $(x,y,z)$ is expressed as
%
\beqa\label{eq_Gauss}
E(x,y,z) \propto
\qquad \qquad \qquad \qquad \qquad \qquad \qquad \qquad \qquad \nnb\\
\frac{w_0}{w(z)}\exp
\left\{
-i[kz-H(z)] - r^2 \left( \frac{1}{{w(z)}^2}+\frac{ik}{2R(z)} \right)
\right\},
\eeqa
%
where $k=2\pi/\lambda$, $r=\sqrt{x^2+y^2}$, $w_0$ is the minimum waist,
which cannot be smaller than $\lambda$ due to the diffraction limit, and
other definitions are as follows:
%
\beqa\label{eq_wz}
{w(z)}^2 = {w_0}^2
\left(
1+\frac{z^2}{{z_R}^2}
\right),
\eeqa
\beqa\label{eq_Rz}
R = z
\left(
1+\frac{{z_R}^2}{z^2}
\right),
\eeqa
\beqa\label{eq_etaz}
H(z) = \tan^{-1}
\left(
\frac{z}{z_R}
\right),
\eeqa
\beqa\label{eq_zr}
z_R \equiv \frac{\pi{w_0}^2}{\lambda}.
\eeqa
Based on the Gaussian laser parameters above,
we now estimate the effective luminosity
${\cal L}$ over the propagation volume of the laser pulse.
We now restore the physical dimensions of $\hbar$ and $c$ in this section,
unless explicitly noted.
Consider a Gaussian laser pulse with duration time $\tau$,
with the speed of light $c$ and an
average number of photons $\bar{N}$ per pulse.
The exchange of a low-mass field may take place anywhere within the volume
defined by the transverse area of the Gaussian laser times the focal length $f$
before reaching the focal point.
We first consider the effective number of photons $N_{int}$ during
an interaction with the time scale of $\Delta t$.
As a result of the interaction we observe a frequency-doubled photon
in the laboratory frame. The momentum transfer of $\sim \hbar\omega/c$
between photons defines the minimum interaction time scale from the
uncertainty principle as follows
\beqa\label{eq_Tint}
\Delta t > 2\pi \omega^{-1}.
\eeqa
The effective number of photons during $\Delta t$ is expressed as
\beqa\label{eq_Nint}
N_{int} = \frac{\Delta t}{\tau}\bar{N}.
\eeqa
Making the pulse duration $\tau \sim \Delta t$
maximizes the instantaneous luminosity.
Suppose a point $z$ along the laser propagation axis.
The instantaneous luminosity at the point $z$ is defined as
\beqa\label{eq_Linstant}
{\cal L}(z) = \frac{C(N_{int},2)}{\pi w^2(z)}
\sim \frac{N^2_{int}}{2\pi w^2_0}\frac{z^2_R}{z^2+z^2_R}
\eeqa
where
$C(N_{int},2)$ denotes a combinatorics to choose two photons amongst
a large number of photons available within the time scale $\Delta t$, and
the expression $w^2(z)$ in Eq.~(\ref{eq_wz}) is substituted to
obtain the second part with the approximation for the combinatorics.
We then consider the averaged instantaneous luminosity $\bar{{\cal L}}$
over the focal length $f$ as follows,
\beqa\label{eq_Lbar}
\bar{{\cal L}} = f^{-1} \int^f_0 {\cal L}(z) dz \sim
\qquad \qquad \qquad \qquad \qquad \nnb\\
\frac{N^2_{int}}{2\pi f w^2_0} z_R \tan^{-1}(f/z_R)
= \frac{N^2_{int}}{2 f \lambda} \tan^{-1}(f/z_R).
\eeqa
The number of effective bunches $b$ is related with $f$ as
\beqa\label{eq_b}
b = \frac{f}{c\Delta t}.
\eeqa
The effective luminosity ${\cal L}$ over the propagation volume of the
laser pulse is finally expressed as
\beqa\label{eq_L}
{\cal L} = b \bar{{\cal L}}
\qquad \qquad \qquad \qquad \quad
\qquad \qquad \qquad \qquad \qquad \nnb\\
= \frac{f}{c\Delta t} \frac{N^2_{int}}{2f\lambda}\tan^{-1}(f/z_R)
= \frac{\Delta t}{\tau} \frac{\bar{N}^2}{2c\tau \lambda}\tan^{-1}(f/z_R),
\eeqa
where Eq.~(\ref{eq_Nint}) is substituted in the last step.

The minimum beam waist $w_0$ at $z=0$ varies with the experimental conditions,
the focal length $f$, and the diameter of the incident beam $d$,
\beq\label{eq_dfw0}
w_0 \sim \frac{2f}{\pi d}\lambda,
\eeq
where the approximation is valid when $f\lambda \ll (\pi/4) d^2$.
The uncertainty on the incident angle between two light waves is 
then expected to be
\beqa\label{eq_Delta}
\Delta\vartheta 
\sim \frac{\lambda}{\pi w_0} \sim \frac{d}{2f}.
\eeqa
We see that $\Delta\vartheta$ is controlled via $w_0$ by choosing
suitable values for $f$ and $d$ in experiments.

The possible uncertainty on the incident angle $\vartheta$ affects
the average of the squared amplitude $\overline{|{\cal M}|^2}$ as shown in
the first part of Eq.~(\ref{eq_the2eps}).
In order to obtain an approximation close enough to reality,
we plug the following step function into Eq.~(\ref{eq_the2eps}):
\beqa\label{eq_rho}
\rho(\vartheta) = \left\{
\begin{array}{ll}
1/\Delta\vartheta & \quad \mbox{for $0 < \vartheta \le \Delta\vartheta$} \\
0 & \quad \mbox{for $\Delta\vartheta < \vartheta \le \pi/2$}
\end{array}
\right\},
\eeqa
which is normalized to the physically possible range $0< \vartheta \le \pi/2$.
By substituting Eq.~(\ref{eq_rho}) into Eq.~(\ref{eq_x2xi}), we obtain
\beqa\label{eq_M2xi}
\overline{|{\cal M}|^2} =
\frac{(4\pi)^2}{2\omega_{opt}^2} \frac{\vartheta_r}{\Delta\vartheta} a
\qquad \qquad \qquad \qquad \qquad \quad \nnb\\
\int^{a^{-1} \{1-(\vartheta_r/\Delta\vartheta)^2 \}}_{-\infty}
\frac{1}{(1-a\xi)^{3/2}} \frac{1}{(\xi^2+1)} d\xi,
\eeqa
where the first factor of the integral kernel corresponds to
the weight function and the second is the Breit-Wigner function (BW) 
with a width of unity.
The weight function of the kernel is close to unity
for small $\xi$, due to the smallness of $a$ in Eq.~(\ref{eq_a}).
Therefore, the value of the integral in Eq.~(\ref{eq_M2xi}) is almost
equivalent to that of BW~\cite{footnoteBW}. This is because
the monotonic positive weight function approaches zero
as $\xi \rightarrow -\infty$ more rapidly than the pure BW,
whereas the pure BW suppresses the increase
of the weight function close to zero
at $\xi \rightarrow a^{-1} \{1-(\vartheta_r/\Delta\vartheta)^2\}$
for $\Delta\vartheta < 1$.
We then approximate Eq.~(\ref{eq_M2xi}) as the integrated BW
over $\pm \sim \infty$ as follows:
\beqa\label{eq_M2final}
\overline{|{\cal M}|^2} \sim
\frac{(4\pi)^2}{2\omega_{opt}^2} \frac{\vartheta_r}{\Delta\vartheta} a \pi.
\eeqa
Let us introduce the integrated form of the cross section in Eq.~(\ref{kinm_13})
over the solid angle $2\pi \sin\theta_3 d\theta_3$ up to
$\overline{\theta_3}$, which is equivalent to the lower limit of the frequency
$\overline{\omega_3}/\omega \equiv 2-h$ though the relation of Eq.~(\ref{kinm_8})
by specifying the frequency deviation $0<h\ll1$ from the exact $2\omega$
as follows
\beqa
\overline{\sigma(h)}=
\frac{\overline{|{\cal M}|^2}}{(8\pi \omega)^{2} \sin^{-4}{\vartheta}}
\int_0^{\overline{\theta_3}} \left( \frac{\omega_3}{\omega} \right)^2 
2\pi\sin\theta_3 d\theta_3\nnb\\
\sim \pi (h/2) (8\pi\omega)^{-2} \vartheta^{-2} |{\cal M}|^2,
\label{eq_IntXsec}
\eeqa
where $\vartheta \ll 1$ is used for the approximation in the last step.
With $a$ in Eq.~(\ref{eq_a}) and $\overline{|{\cal M}|^2}$ in Eq.~(\ref{eq_M2final}),
the weighted cross section in Eq.~(\ref{eq_IntXsec}) is finally expressed as
\beqa\label{eq_dSdO3}
\overline{\sigma(h)}
\sim
\frac{\pi^2}{32} \left( \frac{2\pi}{\lambda} \right)^{-2}
\left(\frac{\vartheta_r}{\Delta\vartheta}\right)
\left(\frac{g m}{M}\right)^2
\vartheta^{-2}_r h,
\eeqa
where the approximation $\vartheta_r \ll 1$ is also taken into account.

Multiplying Eq.~(\ref{eq_L}) by Eq.~(\ref{eq_dSdO3}),
we obtain the integrated yield ${\cal Y}$ above the lowest
frequency $\overline{\omega_3}$ specified by $h$
per laser pulse focusing as follows:
\beqa\label{eq_dYdO3} 
{\cal Y}
= {\cal L} \overline{\sigma(h)}
\quad \qquad \qquad \qquad \qquad \qquad \qquad \qquad \nnb\\
= \frac{1}{256} \frac{\Delta t}{\tau} \frac{\lambda}{c\tau}
\tan^{-1}(f/z_R)
\left(\frac{\vartheta_r}{\Delta\vartheta}\right)
\left(\frac{g m}{M}\right)^2 \vartheta^{-2}_r h \bar{N}^2.
\eeqa

There are several experimental knobs to affect the observable events in
Eq.~(\ref{eq_dYdO3}).
If we choose $\tau \sim \Delta t \sim \lambda/c$, resulting in
$c\tau \sim \lambda$, we can maximize the effective luminosity.
From Eq.~(\ref{eq_Delta}), the reduction of 
$\Delta\vartheta$ or increasing $w_0$
enhances the yield in the case of very low-mass particle exchange. 
From Eq.~(\ref{eq_zr}) and (\ref{eq_dfw0}), we express $f/z_R$ as
\beq\label{eq_f2zr}
f/z_R \sim \frac{\pi d}{4 f \lambda}.
\eeq
From this relation, a shorter focal length enlarges $f/z_R$. 
This introduces a slight increase for the factor $\tan^{-1}(f/z_R)$,
though its effect is tempered by the nature of $\tan^{-1}$.

As a short summary, we make the most important note from the experimental
point of view based on this conceptual design. The condition
$\vartheta_r/\Delta\vartheta = 1$ maximizes the chance
to search for a resonance, while $\vartheta_r/\Delta\vartheta > 1$ results
in a huge suppression of the cross section by $M^{-4}$ ($\hbar=c=1$)
as we discussed.
This is because the resonance peak is out of the region covered
by $\Delta\vartheta$. This parameter corresponds to a sharp cut-off
of the cross section. Therefore, controlling $\Delta\vartheta$ via the relations
of Eq.~(\ref{eq_Delta}), Eq.~(\ref{eq_dfw0}), and Eq.~(\ref{eq_f2zr}) can provide an
experimental way to define the mass range that we exclude,
if no signal is found. From the resonance condition in Eq.~(\ref{eq_CMreso}),
we evaluate the cut-off on the mass range we can exclude by this
conceptual design as follows ($\hbar=c=1$):
\beqa\label{eq_mbound}
m_{cut} \equiv 2\Delta\vartheta\omega_{opt} = \frac{d}{f}\omega_{opt},
\eeqa
where Eq.~(\ref{eq_Delta}) was substituted. 
We note that the measurement is still sensitive to the mass range
even below this cut-off, however, we cannot state where the mass is located
below this limit. Therefore, if we cannot find any symptoms below this
cut-off, we can exclude the entire mass range below this cut-off.
On the other hand, if something is found, we must make effort to
lower the cut-off by introducing a longer focal length and a smaller
beam diameter in order to determine the location of the mass
at which the signal disappears.

\section{Sensitivity enhancement by high-intensity lasers}
   \label{sec4}

As a demonstration we now discuss how much high-intensity lasers improve
the accessibility to the weak coupling domain for the following
reference case, based on a quasi-parallel colliding system.  

First, we note that $\Delta t$ in Eq.~(\ref{eq_Tint}) is the resolvable 
minimum time scale.
As long as we discuss an extremely low-mass field, the interaction
time scale may be over $\hbar/mc^2$.
In the case of low mass below 1~eV, the interaction time scale may be
much longer than $2\pi \omega^{-1}_{opt}$. On the other hand,
photon-photon interactions must occur within a laser pulse duration time $\tau$.
This fact implies that we should assume $\Delta t/ \tau = 1$
even if $\Delta t \gg \tau$.
Since the beam waist of focused laser becomes smaller as it approaches 
the diffraction limit, a short-pulse laser with the duration time close to
$2\pi \omega^{-1}_{opt}$ maximizes the the effective luminosity per laser shot.
Therefore, we assume this condition in the following discussion.

\begin{figure*}
\includegraphics[width=1.0\linewidth]{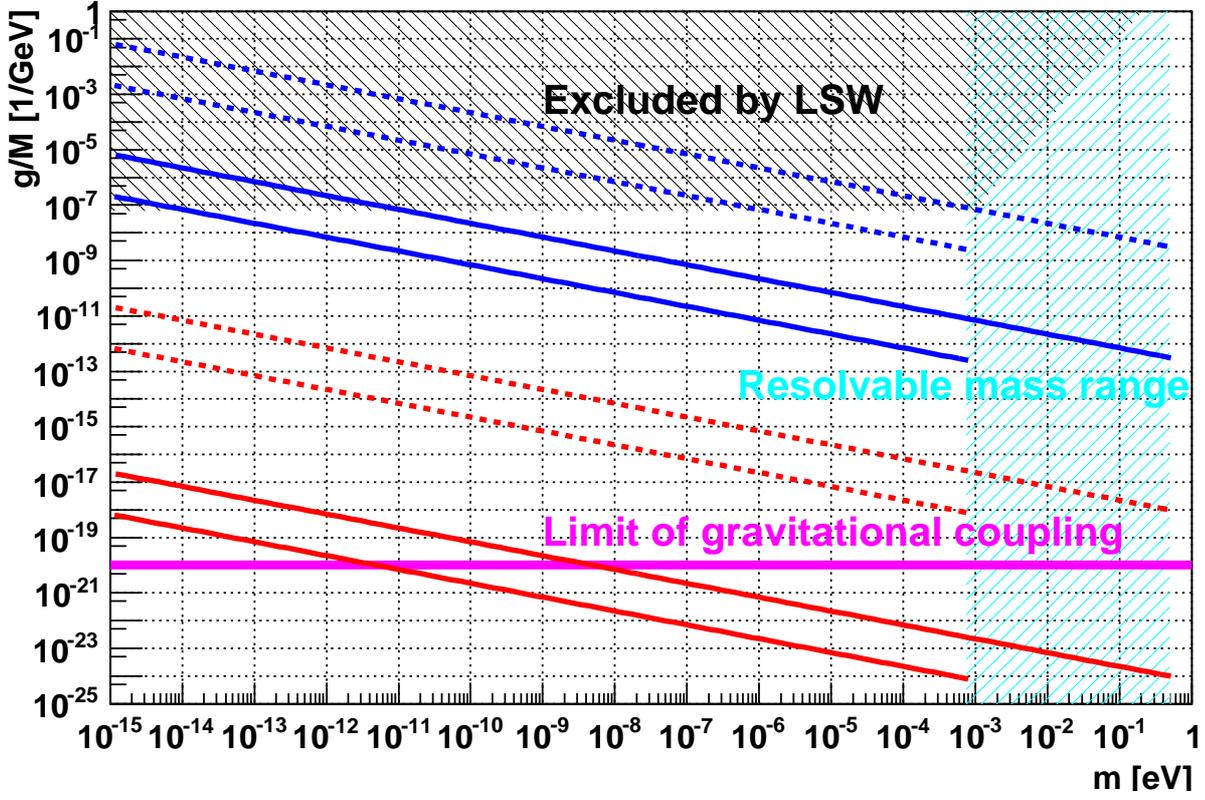}
\caption{
Accessible mass-coupling domain by searching for higher-harmonic generation.
Blue and red lines are cases for spontaneous and induced decays, respectively.
The dotted and solid lines indicate cases for the mean number of photons
per laser pulse $\bar{N} = 2$~J and $\bar{N} = 20$~kJ, respectively. 
The upper and lower lines on each line style
correspond to the case for focal length $f = 1$~m and $f = 1$~km, respectively.
The sky-blue band shows the resolvable mass range.
The shaded area is the excluded domain by laboratory
laser-driven experiments "Light Shining through a Wall (LSW)"\cite{AxionLimit}.
See text for the detail of the other experimental parameters.
}
\label{Fig4}
\end{figure*}

We are now ready to estimate the sensitivity to the coupling $g/M$
for a given average number of photons $\bar{N}$ in the laser pulse,
in the case that ${\cal Y}$ is found to be zero per laser shot based 
on Eq.~\reflef{eq_dYdO3}) for the basic experimental parameters discussed above:
$\omega_{opt} \sim 1$~eV,
$\tau \sim 10$~fs,
$\Delta t/ \tau = 1$,
$d \sim 1$~m,
and $h = 0.1$ to require the frequency close to $2\omega_{opt}$.
Figure \ref{Fig4} shows accessible lower bounds on $g/M$ as a function of $m$
via the search for higher-harmonic generation by focusing a single laser shot.
The blue dotted and solid lines indicate cases for $\bar{N} = 2$~J and 
$\bar{N} = 20$~kJ, respectively. The upper and lower lines on each line style
correspond to the case for $f = 1$~m and $f = 1$~km, respectively.
As we discussed with Eq.(\ref{eq_mbound}), the cut-off values on mass
are lowered by introducing longer focal lengths.
The sky-blue band shows the resolvable mass range by gradually changing
focal lengths from 1~m to 1~km in which we can determine a mass
based on the disappearance of higher harmonic signals from the appearance state.
The black shaded area is the excluded region by conventional
laser-driven experiments so called "Light Shining through a Wall (LSW)"
\cite{AxionLimit}.
In LSW a laser pulse propagates under a static magnetic field $B$ and
a low-mass particle $\sigma$ is assumed to be produced via two photon coupling
to the particle ($\sigma-B-$laser coupling). 
Since $\sigma$ is expected to couple with matter only weakly, 
$\sigma$ can penetrate a wall which prevents penetrations of laser photons 
on the other hand. The particle may couple again to $B$-field placed over 
the wall resulting in a photon in the induced decay process.

%
%
%
%
Our arguments so far have been based on the approach
in which photons both in initial and final states are treated incoherently, 
giving an observable yield proportional to $\bar{N}^2\overline{|{\cal M}|^2}$.  
It is worth noting, however, that the intrinsic nature of laser fields may 
further improve the sensitivity to even smaller couplings
as weak as the gravitational coupling $g/M_P \sim 10^{-20}$GeV${}^{-1}$
with $g\sim\alpha=1/137$ which is indicated by the magenta 
line in Fig.\ref{Fig4}.
Degenerated photons in a laser beam can induce the decay of resonantly produced 
low-mass fields into a degenerated final state as in the case of the
static magnetic field to induce the decay of low-mass particle used for LSW.
If this is the case, compared to their spontaneous decays in vacuum,
we can expect an enhancement factor $\sim N^{1/2}_{ind}$ 
in the scattering amplitude, caused by the creation operator to the degenerated 
state, where $N_{ind}$ is the number of 
degenerated photons involved in the induced laser field~\cite{Induced,DEsearch}.
For example, if we limit $h$ in Eq.(\ref{eq_IntXsec}) to a narrow range 
around $0.5$ in the spontaneous decay process, a phase space decaying into 
$\omega_3 \sim 1.5\omega_{opt}$ and $\omega_4 \sim 0.5\omega_{opt}$ 
can be selected. 
If we induce the $\omega_4 \sim 0.5\omega_{opt}$ by the degenerated state, 
the probability to emit $\omega_3 \sim 1.5\omega_{opt}$ photons is enhanced 
accordingly.
If we rescale all photon frequencies by a factor of two, what is suggested here
is that we simply look for higher-harmonic generation via the following process:
$2\omega + 2\omega \rightarrow 1\omega + 3\omega$, where $1\omega$ is assigned
as the induced field. In this case the yield of the $3\omega$ wave should follow
$\bar{N}^2 N_{ind}\overline{|{\cal M}|^2}$. 
Similar ideas based on the classical treatment are discussed 
in~\cite{Dobrich1,Dobrich2}.
The red lines in Fig.\ref{Fig4} show the lower bounds when $\bar{N} = N_{ind}$
is assumed. The conditions and conventions are same as those for blue lines 
except this additional induced field.
It is a bit of a surprise that it is not impossible
to reach weaker coupling domains even beyond the gravitational coupling 
in a scattering experiment by accumulating statistics over a reasonable 
time period, if we could introduce the induced $1\omega$ waves into 
the same geometry as $2\omega$ to make them focus at a time.
We note that this method is also applicable to nsec pulses, 
typical for MJ-class laser facilities,
as long as the energy per pulse is larger 
than that of short laser pulses
which may compensate the longer time duration 
by the $\bar{N}^2 N_{ind}$ dependence of the higher harmonic yield.
The actual coupling limit must be evaluated, 
eventually based on the statistics of the background of higher harmonics.

%
%
A major instrumental background for the frequency shifted radiation 
is in principle expected to be higher-harmonic generation~(HHG)
from the final focusing optical element and the one to reflect HHG to
the photon detectors. The dominant source of HHG
may be the interface between the residual gas and the surface of the
optical element, where the centrosymmetry is maximally broken.
Even from the maximal estimate $\sim10^{13}$W/cm${}^2$ for a typical damage
threshold which is a much lower intensity compared to HHG due to
relativistic motion of surface electrons, we expect a negligible amount of
$10^{-10}$ for the case of second-harmonic photons from a 1~m${}^2$ aperture
size with a 10fs irradiation, if the optical components are housed
in a vacuum containing $10^{10}$~atoms/cm${}^3$
($\sim 10^{-5}$~Pa)~\cite{SHG}.
The confirmation of a negligible contribution of the background from higher
harmonics is a crucial subject for the present concept.

%
%
As a dominant physical background we expect the lowest-order QED
photon-photon scattering with a forward cross section  $\sim
(\alpha^2/m_e^4)^2 \omega^6 \vartheta^4$~\cite{XsecQED}.
This turns out to be much smaller than Eq.~\reflef{eq_dSdO3}), due to
the specific behavior with respect to the incident angle $\vartheta$.
This indicates that the lowest-order QED contribution is negligible.

If there will be no signal in the single-beam focusing, we only have to
update the condition such that it satisfies $\Delta\vartheta > \vartheta_r$
for heavier masses by increasing $\Delta\vartheta$.
In such a heavier mass region, however, the
two-beam crossing geometry relaxes the constraints on the optical design
such as the focal length.
In either case the single-beam focusing setup considered in this paper
provides a basis to define the mass-coupling limit as well as
the necessary beam intensity as we have demonstrated here.

\begin{figure*}
\includegraphics[width=1.0\linewidth]{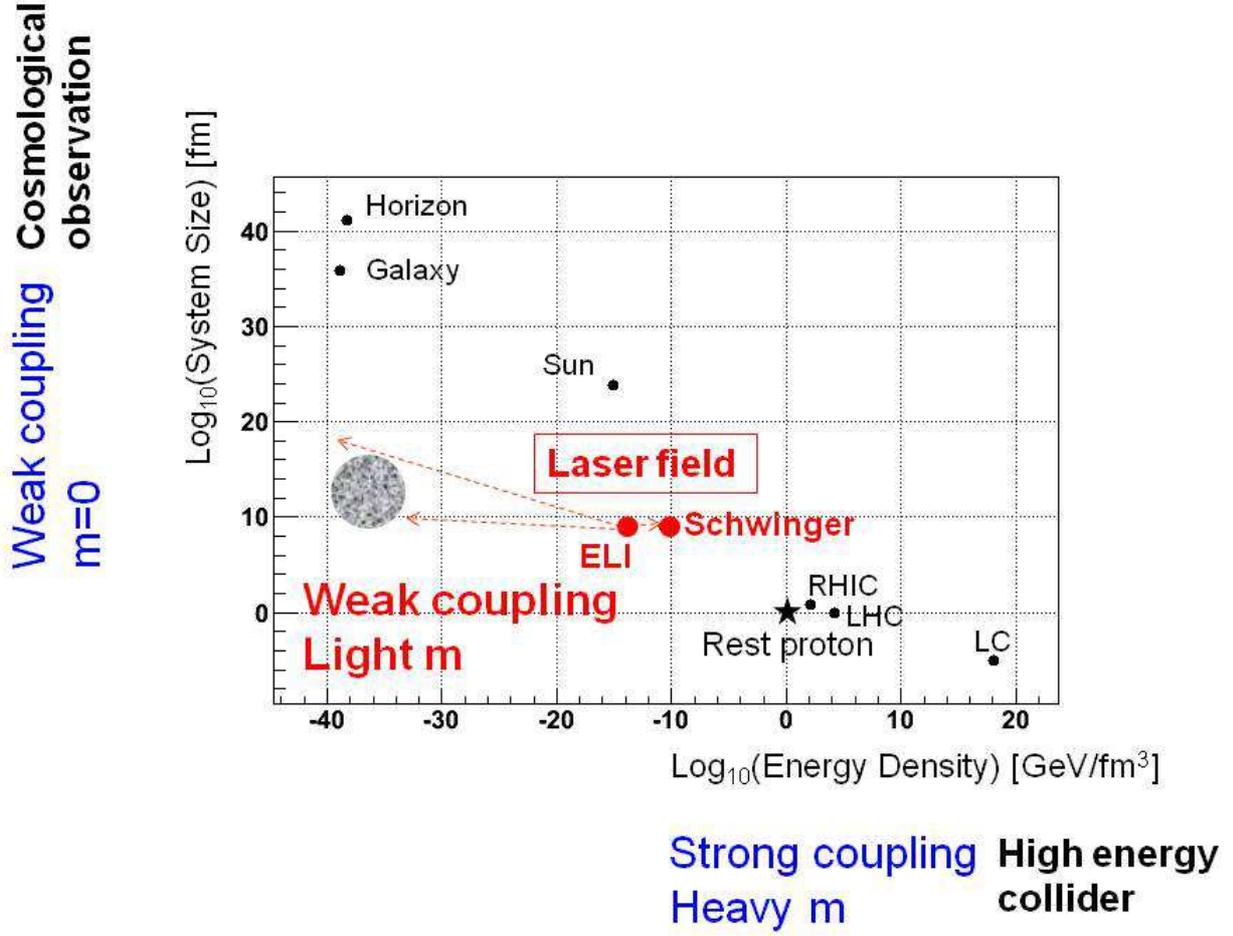}
\caption{
Experimental domains of various approaches to probe matter and vacuum
as a function of the system size vs. the energy density.
Selected systems are
LC: electron-positron collision in the center-of-mass energy $E_{cms}=1$~TeV at
the future linear collider~\cite{Colliders}, assuming
the electron size $10^{-18}$cm which is the upper limit of the electron
radius obtained in high-energy collider experiments;
LHC: proton-proton collision in $E_{cm}=14$~TeV
at the Large Hadron Collider~\cite{Colliders};
RHIC: gold-gold collision in $E_{cm}=200$~GeV per nucleon pair at the
Relativistic Heavy Ion Collider~\cite{Colliders}, the rest proton
indicated by the asterisk as the origin of this plot;
ELI: an optical laser pulse expected in the ELI project~\cite{ELI};
Schwinger: the Schwinger limit~\cite{Schwinger},
Sun, the Milky Way Galaxy and the cosmic horizon with $\Omega_{tot} \sim 1.0$ and
$h \equiv H_0$/100~[km/s/Mpc]$\sim 0.7$~\cite{Cosmology}.
The energy density axis is qualitatively interpreted as the inverse of
the force range or the mass scale $m$ of the exchanged force, because
the mean free path becomes shorter in higher density states, as
long as the coupling to matter is not weak.
On the other hand, the coupling strength to matter $gM^{-1}$ as defined in
Eq.~(\ref{eq_phisigma}) qualitatively reflects the necessary size of matter
or vacuum in order to make the interaction manifest.
The arrow to the higher energy
density towards the Schwinger limit is the direction to probe nonlinear QED
interactions and also towards an understanding of the non-perturbative nature of
the intense field. The arrows directing to the lower energy density region
indicate the extensible domain by using co-propagating intense laser fields,
since the sensitive mass range is below $\sim 1$~eV covering mass scale
relevant for dark energy
and the coupling may be probed to a scale as weak as gravitational coupling
for lighter mass scales. The energy density in this direction depends on
the context. In the context of the scalar field as a candidate of
dark energy in \cite{DEreview,DEsearch},
the energy density should be close to that of the cosmic horizon.
}
\label{Fig9}
\end{figure*}

\section{Conclusion}
   \label{sec5}

We have suggested an approach to probe the nature of vacuum with intense lasers.
The resonance search for extremely light-mass fields
via higher-harmonic generation has been explored by 
focusing high-intensity laser beams.
This is similar to the idea already developed to probe matter.

In this method we take note of the nonlinearities of
vacuum that are
speculated to wait for our sensitive detection of an extraordinarily
feeble signal. In order to detect weak nonlinearities, we need to
spectacularly enhance the signal. The large leap in enhancing these signals is
garnered by the combination of (i) the rapid development of the intense laser
technology and its adoption here; (ii) the employment of our suggested
scheme allows for an enhanced
interaction with the pursued fields. The former element (i) may be brought in,
for example, by an intense optical laser beyond 1~kJ. For the latter factor (ii)
we have suggested a method to include a resonance with
co-propagating photons for the exploration of possible new low-mass fields,
with the aid of induced laser fields to promote the decay into a degenerate
vacuum state.

With the detection of higher-harmonic generation in the co-propagating setup,
we should be able to survey a large sweep of the energy domain
of the intermediating vacuum fields.
If and when we pick up some signal in one particular energy range,
perhaps we can zoom in to this specific energy
(and thus wavelength) of photons by arranging the various knobs, such as the
crossing angle and the (long) beating wavelength of the electromagnetic waves. By
setting up a specific resonance cavity, we may be able to further increase the
sensitivity and more deeply study their properties.

Given a highly intense optical laser beyond $1$~kJ per fs-pulse duration
in the near future, the realization of these suggestions may become
an exciting challenge for future experiments investigating
the physics of the vacuum. We might be able to reach coupling strengths as weak as 
the gravitational strength by utilizing induced laser fields.

Figure~\ref{Fig9} illustrates the experimental domains of various approaches
to probe matter and vacuum, in terms of the system size as a function of
the energy density.
The energy density axis is qualitatively interpreted as the inverse of
the force range or the mass scale $m$ of the exchanged force in Eq.~(\ref{mxelm_7}),
because the mean free path becomes shorter in higher-density states as
long as the coupling to matter is relatively strong.
On the other hand, the coupling to matter $gM^{-1}$ in
Eq.~(\ref{mxelm_7}) or Eq.~(\ref{eq_phisigma})
qualitatively reflects the necessary size of matter or vacuum in order
to make the interaction visible.\\
The Galileo-type telescope observes gravitational
phenomena. These are on the extremely
weak coupling scale of $M_P^{-1}$ with zero mass exchange.
High-energy particle colliders, the Rutherford-type microscopes, focus on
particle generation phenomena. These are due to
strong couplings with heavy mass exchanges within the fm scale.
There is a huge gap between these two approaches. In other words, the region of
weak couplings with finite but light mass exchanges has hardly been probed
so far. It is quite natural to start exploring if there exist
important pieces of the puzzle of nature in these domains.
These explorations might grant us deeper
understanding of the nature of vacuum such as dark energy~\cite{DE}.
The progress of modern physics has been simply driven by
those two types of experimental approaches.
The proposed method with high-intensity lasers
probes the semi-macroscopic vacuum compared to particle physics and
on a much smaller scale of vacuum compared to cosmology.
Provided such semi-macroscopic vacuum scope,
we increase our observational window into a new parameter regime of the vacuum.

\vspace{1cm}

{\bf Acknowledgment}\\
The research has been supported by the DFG Cluster of Excellence MAP
(Munich-Center for Advanced Photonics).
K. Homma appreciate
the support from the Grant-in-Aid for Scientific Research no.21654035
from MEXT of Japan.
T. Tajima is Blaise Pascal Chair Laureate at \'Ecole Normale Sup\'erieure.
We express special thanks to Y. Fujii for his deep discussions with us
and thank P.~Thirolf for his careful reading and
suggestions of our manuscript.


\begin{thebibliography}{99}
\bibitem{PDG}
See section for {\it Axions and other similar particles}
in C. Amsler et al. (Particle Data Group),
Phy. Lett. B{\bf 667}, 1 (2008) and 2009 partial update for the 2010 edition.
%
\bibitem{MCP0}
B.~Holdom,
 {\it Two U(1)'S And Epsilon Charge Shifts},
 Phys. Lett.  B {\bf 166}, 196 (1986).
%
\bibitem{MCP1}
B.~Batell and T.~Gherghetta,
 {\it Localized U(1) gauge fields, millicharged particles, and holography},
Phys. Rev.  D {\bf 73}, 045016 (2006) [arXiv:hep-ph/0512356].
%
\bibitem{MCP2}
S.A.~Abel, J.~Jaeckel, V.V.~Khoze and A.~Ringwald,
  {\it Illuminating the hidden sector of string theory by shining light  through a
        magnetic field},
Phys. Lett.  B {\bf 666}, 66 (2008) [arXiv:hep-ph/0608248].
%
\bibitem{MCP3}
H.~Gies, J.~Jaeckel and A.~Ringwald,
   {\it Polarized light propagating in a magnetic field as a probe of  millicharged
    fermions},
Phys. Rev. Lett.  {\bf 97}, 140402 (2006) [arXiv:hep-ph/0607118].
%
\bibitem{DEreview}
  Y.~F.~Cai, E.~N.~Saridakis, M.~R.~Setare and J.~Q.~Xia,
  {\it Quintom Cosmology: Theoretical implications and observations},
  arXiv:0909.2776 [hep-th];
  S.~Tsujikawa,
  {\it Dark energy: investigation and modeling},
  arXiv:1004.1493 [astro-ph.CO].
%
\bibitem{FujiiScalarTensor}
   Y. Fujii and K. Maeda, {\it The Scalar-Tensor Theory of Gravitation}
  (Cambridge Univ. Press, 2003).
%
\bibitem{Weinberg}
S.~Weinberg,
 {\it A Model Of Leptons},
Phys.\ Rev.\ Lett.\  {\bf 19}, 1264 (1967).
%
%
\bibitem{MourouRMP}G. A. Mourou, T. Tajima, and S. V. Bulanov,
{\it Optics in the Relativistic Regime},
Reviews of Modern Physics {\bf 78} 309 (2006).
%
\bibitem{TajimaHighFieldScience}
Tajima T., Mima K., Baldis H. eds., {\it High Field Science}
(Kluwer Academic/Plenum, New York, 2000).
%
\bibitem{TajimaEPJD}T. Tajima, Eur. Phys. J. D {\bf 55} 519-529 (2009).
%
\bibitem{Beat}
A. T. Forrester, R. A. Gudmundsen, and P. O. Johnson, Phys. Rev. {\bf 99},
1691 (1955);
Y. Minami, T. Yogi and K. Sakai, Phys. Rev. A {\bf 78}, 033822 (2008).
%
\bibitem{Franken}
P. Franken, A. E. Hill, C. W. Peters, and G. Weinreich,
Phys. Rev. Lett. {\bf 7}, 118 (1961).
%
\bibitem{EH}
W.~Heisenberg and H.~Euler,
Z.\ Phys.\  {\bf 98}, 714 (1936)
[arXiv:physics/0605038].
%
\bibitem{Weiscop}
V. Weisskopf, Kong. Dans. Vid. Selsk. Math-fys. Medd. {\bf XIV}, 166 (1936).
%
\bibitem{deVega}
H.J.~de~Vega and N.G. Sanchez, astro-ph/0701212.
%
\bibitem{apb-qed}
In preparation for Applied Physics B.

\bibitem{qed-limit}
K. Homma, {\it Fundamental physics on natures of the macroscopic vacuum 
under high intense electromagnetic fields with accelerators },
AIP Conference Proceedings, Volume {\bf 1153}, 49-60 (2009) 
[arXiv:0911.5701v1 [quant-ph]].

%
\bibitem{BFRT}
  R.~Cameron {\it et al.},
  {\it Search For Nearly Massless, Weakly Coupled Particles By Optical
       Techniques},
  Phys.\ Rev.\  D {\bf 47}, 3707 (1993).
%
%
\bibitem{PVLAS}
  E.~Zavattini {\it et al.}  [PVLAS Collaboration],
  {\it New PVLAS results and limits on magnetically induced optical rotation and
       ellipticity in vacuum},
  Phys.\ Rev.\  D {\bf 77}, 032006 (2008)
  [arXiv:0706.3419 [hep-ex]].
%
\bibitem{BMV}
  C.~Robilliard, R.~Battesti, M.~Fouche, J.~Mauchain, A.~M.~Sautivet, F.~Amiranoff and C.~Rizzo,
  {\it No light shining through a wall},
  Phys.\ Rev.\ Lett.\  {\bf 99}, 190403 (2007)
  [arXiv:0707.1296 [hep-ex]];
  M.~Fouche {\it et al.},
  {\it Search for photon oscillations into massive particles},
  Phys.\ Rev.\  D {\bf 78}, 032013 (2008)
%
\bibitem{ALPS}
  K.~Ehret {\it et al.},
  {\it Production and detection of axion-like particles in a HERA dipole magnet:
       Letter-of-intent for the ALPS experiment},
  arXiv:hep-ex/0702023.
%
\bibitem{LIPSS}
  A.V.~Afanasev, O.K.~Baker and K.W.~McFarlane,
   {\it Production and detection of very light spin-zero bosons at optical
   frequencies},
  arXiv:hep-ph/0605250;
  A.~Afanasev {\it et al.},
  {\it New Experimental limit on Optical Photon Coupling to Neutral, Scalar
       Bosons},
  Phys.\ Rev.\ Lett.\  {\bf 101}, 120401 (2008)
  [arXiv:0806.2631 [hep-ex]].
%
\bibitem{OSQAR}
  P.~Pugnat {\it et al.}  [OSQAR Collaboration],
  {\it First results from the OSQAR photon regeneration experiment: No light
       shining through a wall},
  Phys.\ Rev.\  D {\bf 78}, 092003 (2008)
  [arXiv:0712.3362 [hep-ex]].
%
\bibitem{GammeV}
  A.S.~Chou {\it et al.}  [GammeV (T-969) Collaboration],
  {\it Search for axion-like particles using a variable baseline photon
       regeneration technique},
  Phys.\ Rev.\ Lett.\  {\bf 100}, 080402 (2008)
  [arXiv:0710.3783 [hep-ex]];
  A.S.~Chou {\it et al.}  [GammeV Collaboration],
  {\it A Search for chameleon particles using a photon regeneration technique},
  Phys.\ Rev.\ Lett.\  {\bf 102}, 030402 (2009)
  [arXiv:0806.2438 [hep-ex]].
%
\bibitem{DEsearch}
  Y.~Fujii and K.~Homma,
  {\it An approach toward the laboratory search for the scalar field as a candidate
    of Dark Energy},
  arXiv:1006.1762 [gr-qc].
%
\bibitem{ELI}
http://www.extreme-light-infrastructure.eu/.
See also http://www.eli-np.ro/documents/meeting-10-12march\\
Presentations/11-03-2011/Experiments/Homma-ELI-NP-Buchrest-Mar10-12.pdf
%
\bibitem{BW}
For example, see section for {\it Cross-section formulae for specific processes}
in C. Amsler et al. (Particle Data Group),
Phy. Lett. B{\bf 667}, 1 (2008) and 2009 partial update for the 2010 edition.
%
\bibitem{Yariv}Amnon Yariv, {\it Optical Electronics in Modern Communications}
(Oxford University Press, Inc., Oxford,  1997).
%
\bibitem{footnoteBW}
The complicated analytic solution of the integral
is found. We have checked the behavior of the solution around the width $a$.
%
\bibitem{AxionLimit}
  For example, see Figure 2 and section 4 in
  J.~Jaeckel and A.~Ringwald,
  {\it The Low-Energy Frontier of Particle Physics},
  arXiv:1002.0329 [hep-ph].
%
\bibitem{Induced}
For example see Rodney Loudon,
{\it The Quantum Theory of Light 3rd edition} 
(Oxford University Press, New York, 2000).

\bibitem{Dobrich1}
  B.~Dobrich and H.~Gies,
  {\it Axion-like-particle search with high-intensity lasers},
  JHEP {\bf 1010}, 022 (2010)
  [arXiv:1006.5579 [hep-ph]].

\bibitem{Dobrich2}
  B.~Dobrich and H.~Gies,
  {\it High-Intensity Probes of Axion-Like Particles},
  arXiv:1010.6161 [hep-ph].

%
\bibitem{SHG}
V.G. Bordo, Optics Communications {\bf 132}, 62-72 (1996).
%
\bibitem{XsecQED}See p.183 in W.~Dittrich and H.~Gies,
{\it Probing the Quantum Vacuum} (Springer, Berlin, 2007).
%
\bibitem{Colliders}
See section for  {\it High-energy collider parameters}
in C. Amsler et al. (Particle Data Group),
Phy. Lett. B{\bf 667}, 1 (2008) and 2009 partial update for the 2010 edition.
%
\bibitem{Cosmology}
D.N. Spergel et al., Astrophysical Journal Supplement, {\bf 148}, 175 (2004)
%
\bibitem{Schwinger}
    J. Schwinger, Phys. Rev. {\bf 82}, 664 (1951).
%
\bibitem{DE}
A.G. Riess et al., Astron. J. {\bf 116}, 1009 (1998);
S. Perlmutter et al., Nature {\bf 391}, 51 (1998).
%
\end{thebibliography}
\end{document}